\documentclass{ieeeaccess}
\usepackage{cite}
\usepackage{amsmath,amssymb,amsfonts}
\usepackage{algorithmic}
\usepackage{graphicx}
\usepackage{textcomp}
\usepackage{amsmath} 
\usepackage{amssymb}  
\usepackage{epstopdf} 
\newtheorem{thm}{Theorem}

\newtheorem{prop}{Proposition}
\newtheorem{cor}{Corollary}

\def\BibTeX{{\rm B\kern-.05em{\sc i\kern-.025em b}\kern-.08em
    T\kern-.1667em\lower.7ex\hbox{E}\kern-.125emX}}
\begin{document}
\history{Date of publication xxxx 00, 0000, date of current version xxxx 00, 0000.}
\doi{10.1109/ACCESS.2017.DOI}

\title{Closed-Form Expressions of Ergodic Capacity and MMSE Achievable Sum Rate for MIMO Jacobi and Rayleigh Fading Channels}
\author{\uppercase{Amor Nafkha}\authorrefmark{1}, \IEEEmembership{Senior member, IEEE},
\uppercase{and Nizar Demni\authorrefmark{2}}}
\address[1]{SCEE/IETR, CentraleSup\'elec, Avenue de Boulais, Cesson S\'evign\'e, 35576, France (e-mail:amor.nafkha@centralesupelec.fr).}
\address[2]{IRMAR, Universit\'e de Rennes 1, Campus de Beaulieu, Rennes, 35042, France (e-mail:nizar.demni@univ-rennes1.fr).}
\tfootnote{This work was supported by internal funding from CentraleSup\'elec.}

\markboth
{Amor Nafkha \headeretal: Preparation of Paper for IEEE ACCESS JOURNAL}
{Amor Nafkha \headeretal: Preparation of Paper for IEEE ACCESS JOURNAL}

\corresp{Corresponding author: Amor Nafkha (e-mail: amor.nafkha@centralesupelec.fr).}

\begin{abstract}
Multimode/multicore fibers are expected to provide an attractive solution to overcome the capacity limit of the current optical communication system. In the presence of high cross-talk between modes/cores, the squared singular values of the input/output transfer matrix follow the law of the Jacobi ensemble of random matrices. Assuming that the channel state information is only available at the receiver, we derive a new expression for the ergodic capacity of the MIMO Jacobi fading channel. This expression involves double integrals which can be evaluated easily and efficiently. Moreover, the method used in deriving this expression does not appeal to the classical one-point correlation function of the random matrix model. Using a limiting transition between Jacobi and Laguerre polynomials, we derive a similar formula for the ergodic capacity of the MIMO Rayleigh fading channel. Moreover, we derive a new exact closed-form expressions for the achievable sum rate of MIMO Jacobi and Rayleigh fading channels employing linear minimum mean squared error (MMSE) receivers. The analytical results are compared to the results obtained by Monte Carlo simulations and the related results available in the literature, which shows perfect agreement.
\end{abstract}


\titlepgskip=-15pt

\maketitle

\section{Introduction}
\label{sec:introduction}
\PARstart{T}{o} accommodate the exponential growth of data traffic over the last few years, the space division multiplexing (SDM) based on multi-core optical fiber (MCF) or multi-mode optical fiber (MMF) is expected to overcome the barrier from capacity limit of single-core fiber \cite{SDM,MIMO_tech5,SDMA}. The main challenge in SDM occurs due to in-band cross-talk between multiple parallel transmission channels (cores/modes). This non-negligible cross-talk can be dealt with using multiple-input multiple-output (MIMO) signal processing techniques \cite{MIMO_tech1, MIMO_tech2,MIMO_tech3,MIMO_tech4,MIMO_tech5}. Those techniques are widely used for wireless communication systems and they helped to drastically increase channel capacity. Assuming important cross-talk between cores and/or modes, negligible back-scattering and near-lossless propagation, we can model the transmission optical channel as a random complex unitary matrix \cite{Winzer,DFS,Aris}.

In \cite{DFS}, authors appealed to the Jacobi unitary ensemble (JUE) to establish the propagation channel model for MIMO communications over multi-mode/multi-core optical fibers. The JUE is a matrix-variate analogue of the beta random variable and consists of complex Hermitian random matrices which can be realized at least in two different ways \cite{Col,Meh}. One of them mimics the construction of the beta random variable as a ratio of two independent Gamma random variables: the latter are replaced by two independent complex Hermitian Wishart matrices whose sum is invertible. Otherwise, we can draw a Haar-distributed unitary matrix then take the square of the radial part of an upper left corner \cite{Col}. By a known fact for unitarily invariant-random matrices \cite{Meh}, the average of any symmetric function with respect to the eigenvalues density can be expressed through the one-point correlation function, also known as the single-particle density. In particular, the ergodic capacity of a matrix drawn from the JUE can be represented by an integral where the integrand involves the Christoffel-Darboux kernel associated with Jacobi polynomials (\cite{Meh}, p.384). The drawback of this representation is the dependence of this kernel on the size of the matrix. Indeed, its diagonal is written either as a sum of squares of Jacobi polynomials and the number of terms in this sum equals the size of the matrix least one, or by means of the Christoffel-Darboux formula as a difference of the product of two Jacobi polynomials whose degrees depend on the size of the matrix. To the best of our knowledge, this is the first study that derives exact expression of the ergodic capacity as a double integral over a suitable region.

In this paper, we provide a new expression for the ergodic capacity of the MIMO Jacobi fading channel relying this time on the formula derived in \cite{CDLV} for the moments of the eigenvalues density of the Jacobi random matrix. The obtained expression shows that the ergodic capacity is an average of some function over the signal-to-noise ratio, and it has the merit to have a simple dependence on the size of the matrix which allows for easier and more precise numerical simulations. By a limiting transition between Jacobi and Laguerre polynomials \cite{Ism}, we derive a similar expression for the ergodic capacity of the MIMO Rayleigh fading channel \cite{telatar}. Using the derived expressions and the work of McKay \headeretal \cite{McKay2010}, we are able to derive closed-form formulas for the achievable sum-rate of MIMO Jacobi and Rayleigh fading channels employing linear minimum mean squared error receivers. 

The paper is organized as follows. In Section~\ref{sec:notation}, we settle down some notations and recall the definitions of random matrices and special functions occurring in the remainder of the paper. Section~\ref{sec:model} introduces the MIMO Jacobi fading channel and the discrete-time input-output relation. In Section~\ref{sec:EC}, an exact closed-form expression is derived for the ergodic capacity of MIMO Jacobi fading channel. Using the results of the previous section, we derive a new exact closed-form expression of the ergodic capacity of the MIMO Rayleigh fading channel in Section~\ref{sec:Jacobi-wishart}. In both MIMO Jacobi and Rayleigh fading channels, we provide new closed-form expressions for the achievable sum rate of MIMO MMSE receivers in Section~\ref{mmse}. In Section \ref{sec:numerical}, we demonstrate the accuracy of the analytical expressions through Monte Carlo simulations. Finally, Section \ref{sec:conclusions} is devoted to concluding remarks, while mathematical proofs are deferred to the appendices.

\section{Basic definitions and notations}
\label{sec:notation}
Throughout this paper, the following notations and definitions are used. We start with those concerned with special functions for which the reader is referred to the original book of Ismail \cite{Ism}. The Pochhammer symbol $(x)_k$ with $x \in \mathbb{R}$ and $k \in \mathbb{N}$ is defined by  
\begin{equation}
(x)_k = x(x+1)\dots(x+k-1); \; (x)_0 =1
\label{Pochhammer}
\end{equation}
For $x > 0$, it is clear that 
\begin{equation}
(x)_k = \frac{\Gamma(x+k)}{\Gamma(x)}
\end{equation}
where $\Gamma(.)$ is the Gamma function. Note that if $x = -q$ is a non positive integer then 
\begin{equation}
(-q)_k = \left\{
  \begin{array}{l l}
    (-1)^k \frac{q!}{(q-k)!} & \mbox{if} \  k \geq q \\
    0                        & \mbox{if} \  k < q
  \end{array} \right.
\end{equation}

The Gauss hypergeometric function ${}_2F_1(.)$ is defined for complex $|z| < 1$ by the following convergent power series 
\begin{equation}
{}_2F_1(\theta,\sigma, \gamma, z) = \sum_{k = 0}^{\infty}\frac{(\theta)_k(\sigma)_k}{(\gamma)_k k!} z^k
\label{2f1}
\end{equation}
where $(.)_k$ denotes the Pochhammer symbol defined in \eqref{Pochhammer} and $\theta, \sigma, \gamma$ are real parameters with $\gamma \neq \{0, -1, -2,\dots\}$. The function ${}_2F_1(.)$ has an analytic continuation to the complex plane cut along the half-line $[1,\infty[$. In particular, the Jacobi polynomials $P_q^{\alpha,\beta}(x)$ of degree $q$ and parameters $\alpha> -1$, $\beta > -1$ can also be expressed in terms of the Gauss hypergeometric function \eqref{2f1} as follows 
\begin{equation}
P_q^{\alpha, \beta}(x) = \frac{(\eta)_q}{q!} {}_2F_1(-q,q+\eta+\beta,\eta;\frac{1-x}{2})
\end{equation}
where $\eta = \alpha + 1$. An important asymptotic property of the Jacobi polynomial is the fact that it can be reduced to the $q$-th Laguerre polynomial of parameter $\alpha$ through the following limit 
\begin{equation}
\label{LimTra}
L_q^{\alpha}(x) = \lim_{\beta \rightarrow \infty} P_q^{\alpha,\beta}\left(1-\frac{2x}{\beta}\right),\, x >0
\end{equation}

Now, we come to the notations and the definitions related with random matrices, and refer the reader to \cite{Col, P.J.Forrester_book, Meh}. Firstly, the Hermitian transpose and the determinant of a complex matrix $\textbf{A}$ are denoted by $\textbf{A}^\dag$ and $\det(\textbf{A})$ respectively. Secondly, the Laguerre unitary ensemble (LUE) is formed out of non negative definite matrices $\textbf{A}^\dag\textbf{A}$ where $\textbf{A}$ is a rectangular $m \times n$ matrix, with $m \geq n$, whose entries are complex independent Gaussian random variables. A matrix from the LUE is often referred to as a complex Wishart matrix and $(m ,n)$ are its degrees of freedom and its size respectively. Finally, let $\textbf{X} = \textbf{A}^\dag \textbf{A}$ and $\textbf{Y} = \textbf{B}^\dag \textbf{B}$ be two independent $(m_1,n)$ and $(m_2,n)$ complex Wishart matrices. Assume $m_1+ m_2\geq n$, then $\textbf{X}+\textbf{Y}$ is positive definite and the random matrix $\textbf{J}$, defined as $ \textbf{J} = (\textbf{X}+\textbf{Y})^{-1/2}\textbf{X}(\textbf{X}+\textbf{Y})^{-1/2}$, belongs to the JUE. The matrix $\textbf{J}$ is unitarily-invariant and satisfies $\textbf{0}_n \leq \textbf{J} \leq \textbf{I}_{n}$ where $\textbf{0}_n, \textbf{I}_{n}$ stand for the null and the identity matrices respectively\footnote{For two square matrices $A$ and $B$, we write $A \leq B$ when $B-A$ is a non negative matrix.}. If $m_1, m_2 \geq n$ then the matrix $\textbf{J}$ and the matrix $(\textbf{I}_{n} - \textbf{J})$ are positive definite and the joint distribution of the ordered eigenvalues of $\textbf{J}$ has a probability density function given by 
\begin{align}
 \mathcal{F}_{a,b,n}(\lambda_1,...,\lambda_{n})  & = Z_{a,b,n}^{-1} \prod_{1 \leq j \leq n} \lambda_j^{a-1} (1-\lambda_j)^{b-1}  \nonumber \\ & 
  \times [V(\lambda_1, \dots, \lambda_n)]^2 {\bf 1}_{0 < \lambda_1 < \dots < \lambda_n < 1}
\label{eq:tmpconc}   
\end{align}
with respect to Lebesgue measure $d\lambda = d\lambda_1\dots d\lambda_n$. Here, $a = m_1-n+1, b = m_2-n+1$, $Z_{a,b,n}$ is a normalization constant read off from the Selberg integral \cite{CDLV,P.J.Forrester_book}:
\begin{equation*}
Z_{a,b,n} = \prod_{j=1}^{n}  \frac{\Gamma(a+j-1)\Gamma(b+j-1)\Gamma(1+j)}{\Gamma(a+b+n+j-2)}
\end{equation*}
, and $ V(\lambda_1, \dots, \lambda_n) = \prod_{1 \leq j<k \leq n} \left(\lambda_j-\lambda_k \right)$
is the Vandermonde polynomial. Another construction of matrices from the JUE is as follows \cite{Col}: let $\textbf{U}$ be an $m \times m$ Haar-distributed unitary matrix. Let $t$ and $r$ be two positive integers such that $t+r \leq m$ and $t \leq r$. Let also $\textbf{H}$ be the $r \times t$ upper-left corner of $\textbf{U}$, then the joint distribution of the ordered eigenvalues of $\textbf{H}^{\dag}\textbf{H}$ is given by \eqref{eq:tmpconc} with parameters $a=r-t+1$, $b=m-r-t+1$, and $n=t$. The symbol $\mathbb{E}_{\textbf{\nu}}\left[.\right]$ stands for the expectation operation with respect to the random variable $\textbf{\nu}$, the matrix determinant is denoted by $\det\left(.\right)$, and the matrix inverse will be denoted by $\left[.\right]^{-1}$. The $(i,j)$-th element of a matrix $\textbf{A}$ is indicated by $\left[\textbf{A}\right]_{i,j}$.

\section{System Model}
\label{sec:model}
We consider an optical space division multiplexing where the multiple channels correspond to the number of excited modes/cores within the optical fiber. The coupling between different modes and/or cores can be described by scattering matrix formalism \cite{Aris,Beenakker,Forrester2006}. In this paper, we consider $m$-channel lossless optical fiber with $t \leq m $ transmitting excited channels and $r \leq m$ receiving channels. The scattering matrix formalism can describe very simply the propagation through the fiber using $2m \times 2m$ scattering matrix $\textbf{S}$ given as  
\begin{equation}
\textbf{S}=\begin{bmatrix}
{\textbf{R}_{1}} &{\textbf{T}_{2}}\\ 
{\textbf{T}_{1}} &{\textbf{R}_{2}}
\end{bmatrix},
\label{Scattering_matrix_Aris}
\end{equation}
where the $m \times m$ block matrices $\textbf{R}_{1}$ and $\textbf{R}_{2}$ describe the reflection from left to left and from right to right of the fiber, respectively, and ${\textbf{T}_{1}}$ and ${\textbf{T}_{2}}$ describe the transmission through the fiber from left to right and from right to left, respectively. Since the fiber is assumed to be lossless and time-reversal, the scattering matrix must be a complex unitary symmetric matrix, ($i.e.$ $\textbf{S}^\dag \textbf{S}= \textbf{I}_{2m}$). Therefore, the four Hermitian matrices $\textbf{T}_{1}\textbf{T}_{1}^\dag$, $\textbf{T}_{2}\textbf{T}_{2}^\dag$, $\textbf{I}_{m}-\textbf{R}_{2}\textbf{R}_{2}^\dag$, and $\textbf{I}_{m}-\textbf{R}_{1}\textbf{R}_{1}^\dag$ have the same set of eigenvalues $\lambda_1, \lambda_2, . . .,\lambda_m$. Each of these $m$ transmission eigenvalues is a real number belong to the interval $[0,1]$. Assuming a unitary coupling among all transmission modes the overall transfer matrix $\textbf{T}_{1}$ can be described by a $m \times m$ unitary matrix, where each matrix entry $[\textbf{T}_{1}]_{ij}$ represents the complex path gain from transmitted mode $i$ to received mode $j$. Moreover, the transmission matrix $\textbf{T}_{1}$ is Haar distributed over the group of complex unitary matrices \cite{DFS,Aris}. Given the fact that only $t\leq m$ and $r\leq m$ modes are addressed by the transmitter and receiver, respectively, the effective transmission channel matrix $\textbf{H} \in \mathbb{C}^{r \times t}$ is a truncated\footnote{Without loss of generality, the effective transmission channel matrix $\textbf{H}$ is the $r \times t$ upper-left corner of the transmission matrix  $\textbf{T}_{1}$ \cite{Col,Jiang2009}} version of $\textbf{T}_{1}$. As a result, the corresponding MIMO channel for this system reads
\begin{equation}
\textbf{y}=  \textbf{H} \textbf{x} + \textbf{z}
\label{MIMO_input_output_relation}
\end{equation}
where $\textbf{y} \in \mathbb{C}^{r\times 1}$ is the received signal vector of dimension $r \times 1$, $\textbf{x} \in \mathbb{C}^{t\times 1}$ is a $t \times 1$ transmitted signal vector with covariance matrix equal to $\frac{\mathcal{P}}{t}\textbf{I}_{t}$, and $\textbf{z} \in \mathbb{C}^{r \times 1}$ is a $r \times 1$ zero mean additive white circularly symmetric complex Gaussian noise vector with covariance matrix equal to $\sigma^2\textbf{I}_{r}$. The variable $\mathcal{P}$ is the total transmit power across the $t$ modes/cores, and $\sigma^2$ is the Gaussian noise variance.
\section{Ergodic Capacity of MIMO Jacobi channel}
\label{sec:EC}
The expression of the ergodic capacity of the MIMO Jacobi fading channel was firstly expressed in \cite{DFS} as an integral and sum of Jacobi polynomials using the same theoretical approach as in Telatar's paper \cite{telatar}. Recently, ergodic capacity bounds (upper and lower) of the MIMO Jacobi fading channel were derived in \cite{bonnefoi} and \cite{nafkha}. 

In this section, we provide a novel and simple closed-form expression of the ergodic capacity in the setting of MIMO Jacobi fading channel. We assume that the channel state information (CSI) is only known at the receiver, not at the transmitter. The channel ergodic capacity, under a total average transmit power constraint, is then achieved by taking $\textbf{x}$ as a vector of zero-mean circularly symmetric complex Gaussian components with covariance matrix $\mathcal{P}\textbf{I}_{t}/t$, and it is given by \cite[Eq. (10)]{DFS}
\begin{equation}
C_{t,r}^{m,\rho}=\left\{
	\begin{array}{l}
\mathbb{E}_{\textbf{H}}\left[\ln\det\left(\textbf{I}_{t} + \frac{\rho \textbf{H}^\dag\textbf{H}}{t}  \right)\right], t\leq r \\
\mathbb{E}_{\textbf{H}}\left[\ln\det\left(\textbf{I}_{r} + \frac{\rho \textbf{H}\textbf{H}^\dag}{t}  \right)\right], t > r
	\end{array}
\right.
\label{EC_DFS}
\end{equation}
where $\mathbb{E}_{\textbf{H}}[.]$ denotes the expectation over all channel realizations, $\ln$ is the natural logarithm function and $\rho=\frac{\mathcal{P}}{\sigma^2}$ is the average signal-to-noise ratio (SNR). Without loss of generality, we shall assume in the sequel that $t \leq r$ and $m \geq t+r$. Indeed, $\textbf{H}^\dag\textbf{H}$ and $\textbf{H}\textbf{H}^\dag$ share the same non zero eigenvalues while if $m < t + r$, then $(m-r) + (m-t) < m$ and work in \cite[Theorem 2]{DFS} shows that the ergodic capacity is given by
\begin{equation}
C_{t,r}^{m,\rho}=  (t+r-m) C_{1,1}^{1,\rho}+C_{m-r,m-t}^{m,\rho}
\label{m_less_mt_mr}
\end{equation}

In this paper, we assume further that $m > t+r \Leftrightarrow b \geq 2$ and the case $m = r+t \Leftrightarrow b = 1$ can be dealt with by a limiting procedure. Actually, our formula for the ergodic capacity derived below is valid for real $a > 0, b > 1$, and we can consider its limit as $b \rightarrow 1$. However, for ease of reading, we postpone the details of the computations relative to this limiting procedure to a future forthcoming paper. 

Now, recall that the random matrix $\textbf{H}^\dag\textbf{H}$ has the Jacobi distribution, then its ordered eigenvalues have the joint density given by \eqref{eq:tmpconc} with parameters $a=r-t+1$ and $b=m-t-r+1$. Using \eqref{eq:tmpconc}, we can explicitly express the ergodic capacity \eqref{EC_DFS} as
\begin{align}
C_{t,r}^{m,\rho} &= \int_0^1\dots \int_0^1 \sum_{k=1}^{t} \ln\left(1 + \rho \lambda_k \right) \mathcal{F}_{a,b,t}(\lambda_1,\dots,\lambda_{t}) \nonumber \\ & d\lambda_1 \dots d\lambda_{t}
\label{exact_EC_1}
\end{align}

A major step towards our main result is the following proposition.
\begin{prop}
\label{prop1}
For any $\rho \in (0,1)$,
\begin{align}
\Psi C_{t,r}^{m,\rho} &= A_{t,r}\, \rho^{t-1}P_{t-1}^{r-t,m-t-r-1}\left(\frac{\rho+2}{\rho}\right) \nonumber \\ &  {}_2F_1(t+1,r+1,m+1;-\rho)
\end{align}
where the operator $\Psi= \left[D_{\rho}(\rho D_{\rho})\right]$ with $D_{\rho}$ is the derivative operator with respect to $\rho$, and $A_{t,r} =  \frac{r\,t!}{(m-t+1)_{t}}$.
\end{prop}

The full proof for Proposition \ref{prop1} can be found at the Appendix \ref{appen1}. With this proposition in hand, we are able to derive the following new expression of the ergodic capacity under MIMO Jacobi fading channel.
\begin{thm}
\label{new_EC}
For any $\rho \geq 0$, The ergodic capacity of an  uncorrelated MIMO Jacobi fading channel is given by
\begin{align}
 C_{t,r}^{m,\rho} & =B_{t,r}^m \int_{0}^1 u^{a-1}(1-u)^{b-2}P_{t-1}^{a-1,b}(1-2u) \nonumber \\ & \times P_{t}^{a-1,b-2}(1-2u)  Li_2\left(-\rho u \right) du 
\label{new_exp_EC}
\end{align}
where $a=r-t+1$, $b=m-r-t+1$, and $ B_{t,r}^m = \frac{t! \, (m-t)!}{\Gamma(r) \, \Gamma(m-r)}$. The function $Li_2\left(.\right)$ is the dilogarithm function \cite{Morris79} defined as
\begin{equation*}
Li_2\left( z \right) = -\int_{0}^z \frac{\ln(1-u)}{u} du ,\: z \in \mathbb{C}
\end{equation*}
\end{thm}

The appendix \ref{appen2} contains proof of Theorem \ref{new_EC}.

\section{Ergodic capacity of MIMO Rayleigh channel}
\label{sec:Jacobi-wishart}

The ergodic capacity of the MIMO Rayleigh fading channel was extensively examined in order to provide a compact mathematical expression in several papers \cite{telatar,icc2003,oyman,simon,kiessling,maaref}. In \cite{icc2003,oyman}, the ergodic capacity is provided using the Christoffel-Darboux kernel, and the authors replaced the Laguerre polynomials by their expressions which is a known fact in invariant random matrix models. In \cite{simon,kiessling,maaref}, authors derived a closed form expression of moment generating function (MGF) so that the ergodic capacity may be derived by taking the first derivative. However, this expression of MGF relies on the Cauchy-Binet Theorem and only gives a hypergeometric function of matrix arguments \cite{gross}, from which by derivatives, we can get again an alternating sum coming from the determinant. Consequently, we can not derive the new proposed expression of the ergodic capacity \eqref{new_telatar} from this sum.

Using the limiting transition \eqref{LimTra} between Jacobi and Laguerre polynomials, we are able to give another expression for the ergodic capacity expression of the wireless MIMO Rayleigh fading channel. Indeed, it was shown in \cite{Aris,Aris06}, that the parameter $b$ in \eqref{new_exp_EC} can be interpreted as the power loss through the optical fiber. Therefore, as $b$ becomes large, the channel matrix $\textbf{H}$ in \eqref{MIMO_input_output_relation} starts to look like a complex Gaussian matrix with independent and identically distributed entries. As a matter of fact, the MIMO Jacobi fading channel approaches the MIMO Rayleigh fading channel in the large $b$-limit corresponding to infinite power loss through the optical fiber. In particular, the ergodic capacity \eqref{new_exp_EC} converges as $b \rightarrow \infty$ to the ergodic capacity of the uncorrelated MIMO Rayleigh fading channel already considered by Telatar in \cite[Theorem 2]{telatar}, and we are able to derive the following new result. 

\begin{thm}
\label{new_wishart_EC}
The ergodic capacity of the uncorrelated MIMO Rayleigh fading channel with $t$ transmitters and $r$ receivers can be expressed
\begin{align}
 C_{t,r}^{\rho} &= \frac{t!}{(r-1)!} \int_{0}^{+\infty} u^{r-t}\, e^{-u} \, L_{t-1}^{r-t}(u) \, L_{t}^{r-t}(u) \nonumber \\ &  \times  Li_2\left(-\rho u \right) du.
\label{new_telatar}
\end{align}
\end{thm}

The reader can refer to Appendix \ref{appen3} for the proof of Theorem \ref{new_wishart_EC}.

\section{Achievable sum rate of MIMO MMSE receiver}
\label{mmse}
In this section, we are interested in the performance of linear MMSE receivers. Assuming to employ a MMSE filter, and that each filter output is independently decoded. Let $\rho_k$ denotes the instantaneous signal to interference-plus-noise ratio (SINR) to the $k^{th}$ MIMO subchannel\footnote{In our case ($t \leq r$), the MIMO channel can be decomposed into $t$ parallel subchannels.}. Minimizing the mean squared error between the output of a linear MMSE receiver and the actually transmitted symbol $\textbf{x}_k$ for $1 \leq k \leq t$ leads to the filter vector
\begin{equation}
\textbf{g}_k = \left(\textbf{H}\textbf{H}^\dag+\frac{r}{\rho}\textbf{I}_{t}\right)^{-1} \textbf{h}_k
\label{filter_mmse} 
\end{equation}
where $\textbf{h}_k$ is the $k^{th}$ column of channel matrix $\textbf{H}$. Applying this filter vector into \eqref{MIMO_input_output_relation} yields
\begin{equation}
\textbf{x}_k^{mmse} = \textbf{g}_k^\dag \textbf{y}
\label{out_mmse} 
\end{equation}
The achievable ergodic sum rate for the MMSE receiver can be expressed as
\begin{equation}
\mathcal{R} = \sum_{k=1}^{t} \mathbb{E}_{\rho_k} \left[\ln\left(1+\rho_k\right) \right]
\label{rate_mmse} 
\end{equation}
As shown in \cite{verdu1998,McKay2010}, the instantaneous received SINR for the $k^{th}$ MMSE filter output is given by
\begin{equation}
\rho_k= \frac{1}{\left[\left(\textbf{I}_{t} + (\rho/t)\textbf{H}^\dag\textbf{H}\right)^{-1}\right]_{k,k}}-1
\label{sinr_mmse} 
\end{equation}
In general, the analytical closed-form expression of the probability density function of $\rho_k$ is difficult to determine. This situation makes the direct evaluation of the achievable ergodic MMSE sum rate (\ref{rate_mmse}) very difficult.

Let $\textbf{H}_k$ denotes the submatrix obtained by striking $\textbf{h}_k$ out of $\textbf{H}$. As shown in \cite[Theorem 1.33]{Hiai2014}, the $k^{th}$ diagonal term of the matrix $\left(\textbf{I}_{t} + \frac{\rho \textbf{H}^\dag\textbf{H}}{t}\right)^{-1}$ can be expressed as
\begin{equation}
\left[\left(\textbf{I}_{t} + \frac{\rho \textbf{H}^\dag\textbf{H}}{t}\right)^{-1}\right]_{k,k} = \frac{\det\left(\textbf{I}_{t-1} + \frac{\rho \textbf{H}_k^\dag\textbf{H}_k}{t} \right)}{\det\left(\textbf{I}_{t} + \frac{\rho \textbf{H}^\dag\textbf{H}}{t}\right)}
\label{mat_inv}
\end{equation}
the matrix $\textbf{H}_k^\dag\textbf{H}_k$ is the $k \times k$ principal minor of matrix $\textbf{H}^\dag\textbf{H}$ defined by striking out the $k^{th}$ column of $\textbf{H}$.

Similarly to what has been done in \cite{McKay2010}. By using (\ref{sinr_mmse}) and (\ref{mat_inv}) in (\ref{rate_mmse}), we can obtain the following expression of the achievable ergodic sum rate for the MMSE receiver.   
\begin{align}
\mathcal{R} &= t \, \mathbb{E}_{\textbf{H}}\left[\ln\det\left(\textbf{I}_{t} + \frac{\rho \textbf{H}^\dag\textbf{H}}{t}  \right)\right] - \nonumber \\ & \sum_{k=1}^{t} \mathbb{E}_{\textbf{H}_k}\left[\ln\det\left(\textbf{I}_{t-1} + \frac{\rho \textbf{H}_k^\dag\textbf{H}_k}{t}  \right)\right]
\label{rate_mmse_1} 
\end{align}

By employing the Haar invariant property, exchanging any two different rows or/and exchanging two different columns do not change the joint distribution of the entries, the joint probability density function of the ordered eigenvalues of $\textbf{H}_k^\dag\textbf{H}_k$ is the same as $\textbf{H}_j^\dag\textbf{H}_j$ for all $j\neq k$ and $j \in \{1,..,t\}$. Thus, the achievable ergodic sum rate for the MMSE receiver can be expressed as 
\begin{align}
 \mathcal{R} &= t \, \mathbb{E}_{\textbf{H}}\left[\ln\det\left(\textbf{I}_{t} + \frac{\rho \textbf{H}^\dag\textbf{H}}{t} \right)\right] - \nonumber \\ & t \, \mathbb{E}_{\textbf{H}_1}\left[\ln\det\left(\textbf{I}_{t-1} + \frac{\rho \textbf{H}_{t}^\dag\textbf{H}_{t}}{t}\right)\right]
\label{rate_mmse_final} 
\end{align}

In case of MIMO Jacobi fading channel, the matrix $\textbf{H}_{t}$ is the $r \times (t-1)$ left corner of the channel matrix $\textbf{H}$, then the joint distribution of the ordered eigenvalues of $\textbf{H}_{t}^\dag\textbf{H}_{t}$  is given by (\ref{eq:tmpconc}) with parameters $a= r-t+2$, $b=m-r-t+2$, and $n=t-1$. The following result characterizes the achievable ergodic sum rate of the MIMO Jacobi fading channel when the linear MMSE filter is used at the receiver side.

\begin{thm}
\label{new_jaco_MMSE}
For any $\rho \geq 0$, The achievable ergodic sum rate of MMSE receiver under MIMO Jacobi fading channel is given by
\begin{align}
\mathcal{R}_{t,r}^{m,\rho}= t \, \left[ C_{t,r}^{m,\rho} - C_{t-1,m_r}^{m,\frac{(t-1)\rho}{m_t}}\right]
\label{new_Jac_Rate}
\end{align}
\end{thm}

Very recently, Lim \textit{et al.} \cite{lim} proposed closed form expression of the achievable sum rate for MMSE MIMO systems in uncorrelated Rayleigh environments. However, their derived expression, \cite[eq.(67)]{lim}, is not closed form and does not allow a better understand of the MMSE achievable sum rate due the use of the sum of Meijer G-functions (or equivalent representation in terms of  generalized hypergeometric functions). In following corollary, we presented a novel and exact closed-form formula for ergodic achievable sum rate for MMSE receiver under MIMO Rayleigh fading channels. 
\begin{cor}
\label{new_Gau_MMSE}
For any $\rho \geq 0$, The achievable ergodic sum rate of MMSE receiver under MIMO Rayleigh fading channels can be expressed as
\begin{align}
\mathcal{R}_{t,r}^{\rho} &= r \left[\Psi(t,r,\rho) - \Psi(t,r+1,\rho) \right] + \frac{t!}{(r-1)!} \nonumber \\ &  \times \int_{0}^{+\infty} u^{r-t+1}  e^{-u} L_{t-2}^{r-t+1}(u) L_{t-1}^{r-t+1}(u) \nonumber \\& \times \left[Li_2\left(\frac{-\rho u}{t}\right)-Li_2\left(
\frac{-\rho (t-1) u}{t^2}\right)\right] du.
\label{new_Gau_Rate}
\end{align}
where 
\begin{align*}
	\Psi(t,r,\rho) &=  \frac{t!}{(r-1)!} \int_{0}^{+\infty} u^{r-t} e^{-u} \left[L_{t-1}^{r-t}(u) \right]^2 \nonumber \\ &  \times  Li_2\left(\frac{-\rho u}{t}\right) du
\end{align*}
\end{cor}

\section{Numerical Results}
\label{sec:numerical}
\begin{figure}[ht]
\centering
\includegraphics[width=3.2in]{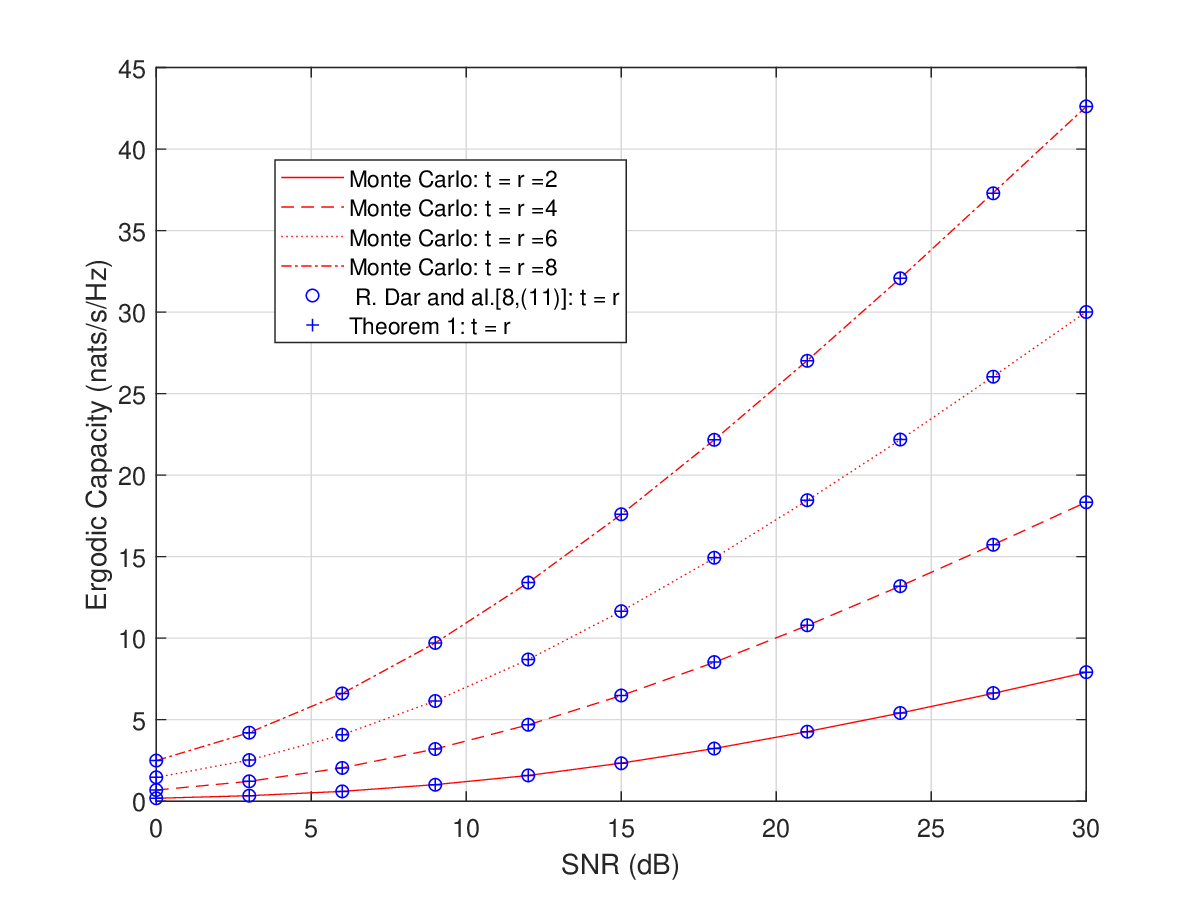}
\caption{The variation of the ergodic capacity of MIMO Jacobi channel as a function of $\rho$ for $m=20$}
\label{fig:accurate1}
\end{figure}

In this section, we present numerical results supporting the analytical expressions derived in Section \ref{sec:EC} and Section \ref{sec:Jacobi-wishart}. All Monte Carlo simulation results are obtained with $10^5$ runs. Herein, we consider the case where the channel state information is available at the receiver side. Figure \ref{fig:accurate1} examines the ergodic capacity of the MIMO Jacobi fading channel as a function of the SNR, when the number of parallel transmission paths is fixed to $m = 20$ and the number of transmit modes equal to the number of receive modes $r=t$. It is evident that when we increase the number of transmitted and received modes, we improve the ergodic capacity of the system. As expected, the ergodic capacity increases with SNR. Figure \ref{fig:accurate1} is also shown that the two theoretical expressions curves of the ergodic capacity \eqref{new_exp_EC} and \cite[(11)]{DFS} perfectly matched the simulation results.

\begin{figure}[ht]
\centering
\includegraphics[width=3.2in]{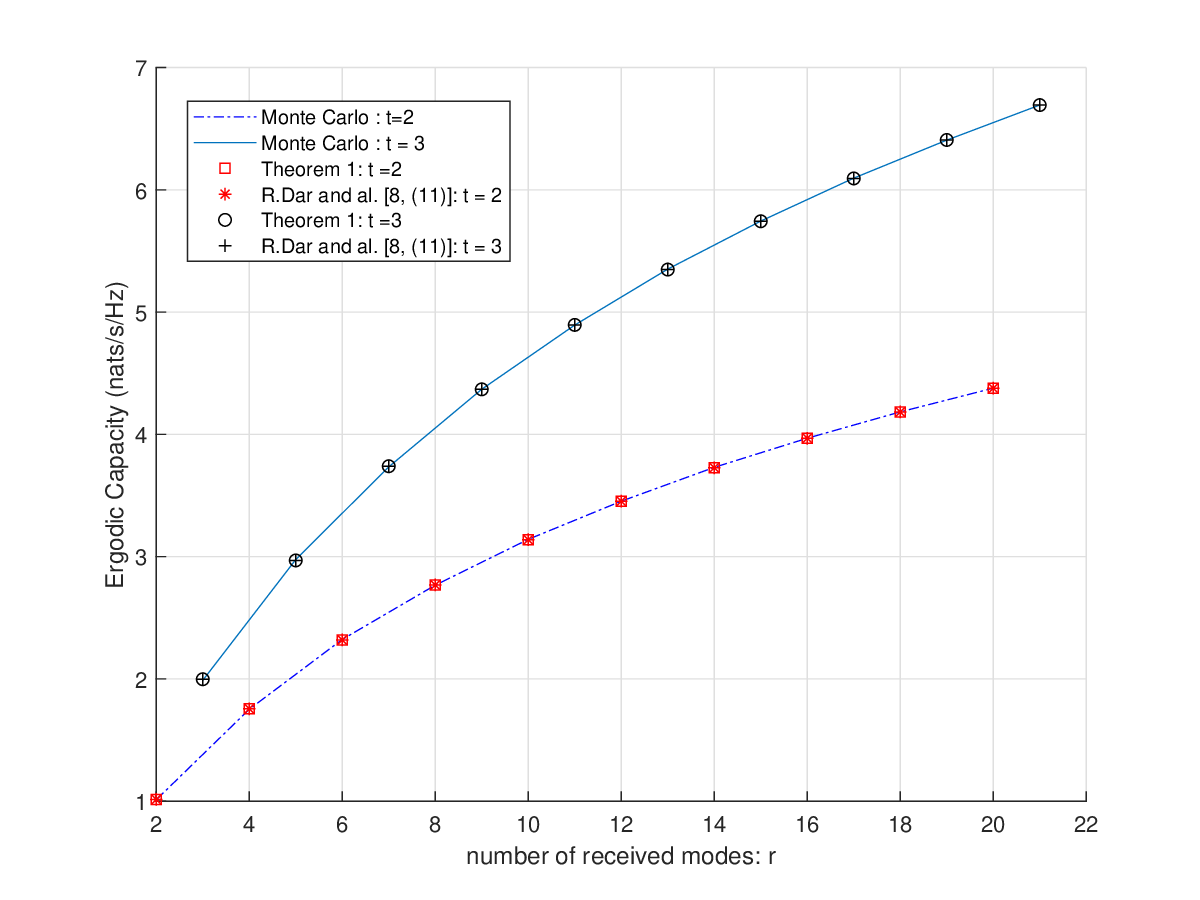}
\caption{Ergodic capacity of MIMO Jacobi channel for various numbers of transmit of receive modes, $\rho=10$dB, and $m=25$.}
\label{fig:accurate2}
\end{figure}

\begin{figure}[ht]
\centering
\includegraphics[width=3.2in]{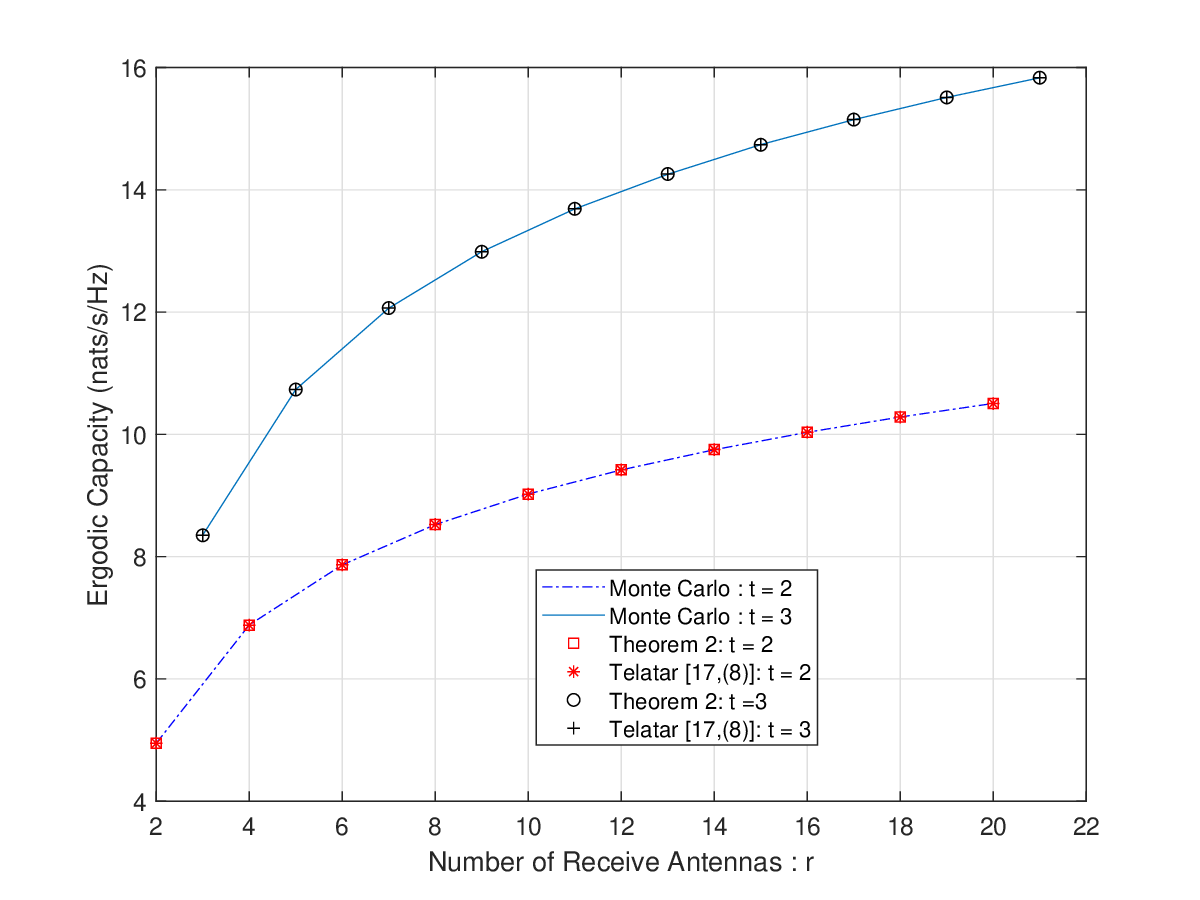}
\caption{Evolution of the capacity versus the number of receiving antennas for $\rho=10dB$}
\label{fig:accurate4}
\end{figure}

Figure \ref{fig:accurate2} shows the theoretical and simulated ergodic capacity of MIMO Jacobi channel as a function of the number of received modes. Here, we fixed the number of parallel transmission paths to $m=25$, the SNR to $\rho = 10$ dB, and the number of transmit modes $t$ to have following values $\{2,3\}$. It is shown that every simulated curve is in excellent agreement with the theoretical curves calculated from \eqref{new_exp_EC} and \cite[(11)]{DFS}. The relationship between the channel capacity and the number of received modes is logarithmic. This implies that trying to improve the channel capacity by just increasing the number of received modes is not efficient. The same relationship has been already noted and discussed in the case of the uncorrelated MIMO Rayleigh fading channel (see Fig. \ref{fig:accurate4} and \cite{telatar}).

\begin{figure}[ht]
\centering
\includegraphics[width=3.2in]{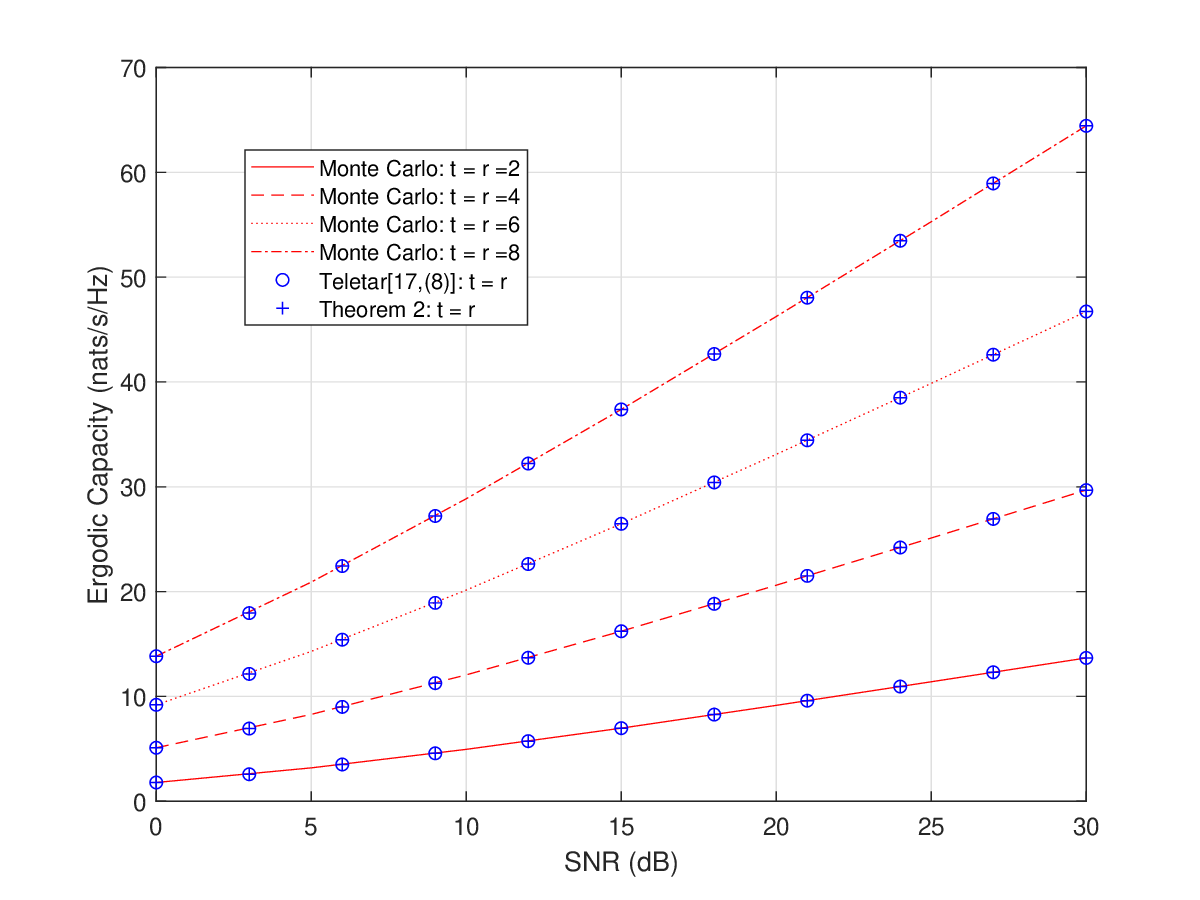}
\caption{Ergodic capacity of the uncorrelated MIMO Rayleigh fading channel versus SNR for different numbers of transmit and receive antennas.}
\label{fig:accurate3}
\end{figure}

For the uncorrelated MIMO Rayleigh fading channel, the proposed expression of the ergodic capacity was verified through Monte Carlo experiments and it is shown in Fig. \ref{fig:accurate3}. We can observe that the expression in \eqref{new_telatar} matches perfectly with the expression introduced by Telatar \cite[Eq. (8)]{telatar}. In Fig. \ref{fig:accurate3}, the comparisons are shown between theoretical expressions and simulation values of the ergodic capacity as a function of the SNR. As we can observe in Fig. \ref{fig:accurate3}, for a given SNR, the capacity increases as the numbers of transmit and receive antennas grow. 

In Fig. \ref{fig:accurate4}, we show the effects of the number of receive antenna elements on the ergodic capacity of uncorrelated MIMO Rayleigh fading channel. As expected, we observe the ergodic capacity increases in logarithm way with increasing number of receive antennas. As for optical MIMO channel, the three different ways to compute the uncorrelated MIMO Rayleigh fading channel capacity give the same results. These simulations were carried out to verify the mathematical derivation and no inconsistencies were noted.

\begin{figure}[ht]
\centering
\includegraphics[width=3.2in]{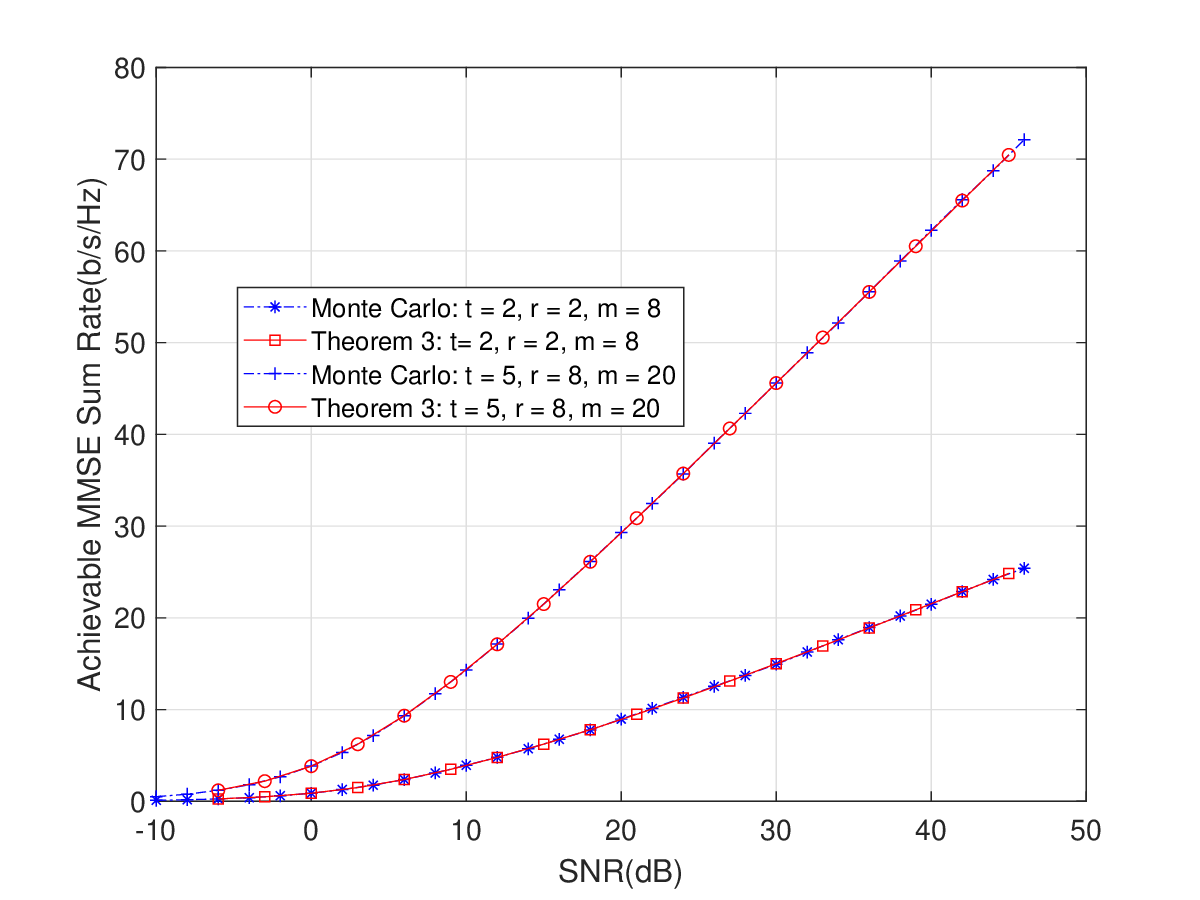}
\caption{Evolution of the ergodic sum rate for the MMSE receiver over MIMO Jacobi fading channel.}
\label{fig:MMSE2}
\end{figure}

We now focus on the ergodic sum rate for the MMSE receiver. We first consider the MIMO Jacobi fading channel. Fig. \ref{fig:MMSE2} shows the evolution of the ergodic sum rate for the MMSE receiver versus the SNR over the optical MIMO channel. For these results, we suppose that either $m=20$ or $m=8$. As expected, the ergodic sum rate increases with increasing SNR. Moreover, our simulation results show that the formula derived in Theorem \ref{new_jaco_MMSE} and Monte Carlo  simulations provide the same results. 

\begin{figure}[ht]
\centering
\includegraphics[width=3.2in]{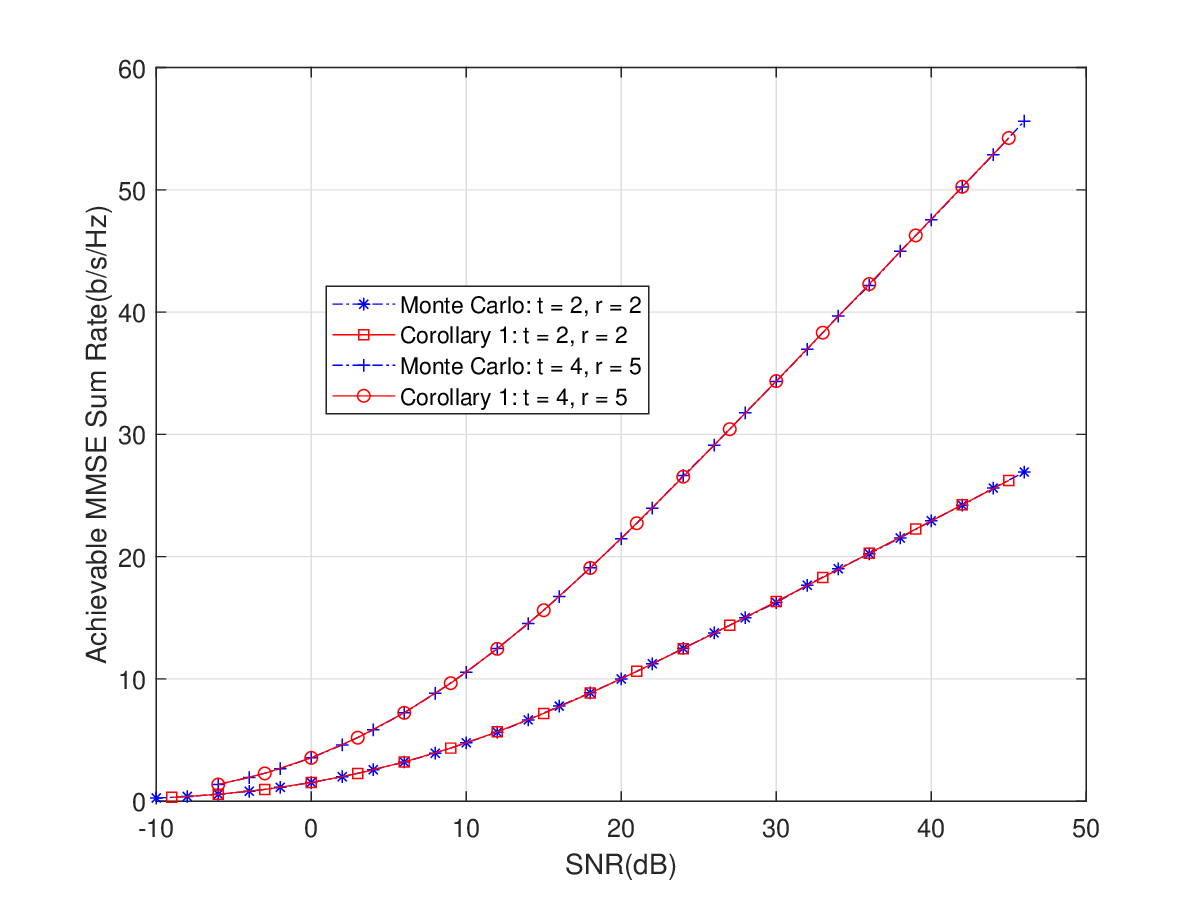}
\caption{Evolution of the ergodic sum rate for the MMSE receiver over an uncorrelated MIMO Rayleigh fading channel.}
\label{fig:MMMSE1}
\end{figure}

Finally, Fig. \ref{fig:MMMSE1} shows the evolution of the ergodic sum rate for the MMSE receiver versus the SNR over an uncorrelated MIMO Rayleigh fading channel. We compare the sum rate obtained by means of Monte Carlo simulations and the one obtained with the formula derived in Corollary \ref{new_Gau_MMSE}. Fig. \ref{fig:MMMSE1} shows that the same results are obtained with both approaches. Moreover, it worth noting that, in the case where $r = t = 2$, the obtained results are the same as in \cite{McKay2010}.

\section{Conclusions}
\label{sec:conclusions}
This paper focused on the MIMO Jacobi fading channel and a new expression is proposed for the ergodic capacity. This new expression allows to better understand the role of each of the parameters and, with this formula, numerical evaluation of the capacity does not require the computation of a sum of Jacobi polynomials. This expression was used to make the connection between optical MIMO channel and wireless MIMO channels and allowed us to propose a new expression for the capacity of the MIMO Rayleigh fading channel. Furthermore, we use the above capacity formulas to find new expressions of the ergodic sum rate with MMSE receiver over both MIMO Jacobi and Rayleigh fading channels. Finally, numerical simulations were used to verify mathematical derivations and show the evolution of the ergodic capacity versus SNR and versus the number of transmitters and receivers.
 \appendices
\section{Proof of Proposition 1}
 \label{appen1}
For ease of reading, we simply denote below the ergodic capacity by $C(\rho)$ and write $n$ for the number of transmitters $t$. Moreover, the reader can easily check that our computations are valid for real $a > 0, b > 1$.

We start by recalling from \cite[Corollary 2.3]{CDLV} that for any $k \geq 1$,
\begin{align*}
\int \left(\sum_{i=1}^n\lambda_i^k\right) \mathcal{F}_{a,b,n}(\lambda)d\lambda &= \frac{1}{k!}\sum_{i=0}^{k-1}(-1)^i\binom{k-1}{i} \nonumber \\ & \prod_{j=-i}^{k-i-1}\frac{(n+j)(a+n+j-1)}{(a+b+2n+j-2)}.
\end{align*}
Now, let $\rho \in [0,1]$ and use the Taylor expansion 
\begin{equation*}
\ln(1+\rho \lambda_i) = \sum_{k=1}^{\infty}(-1)^{k-1}\frac{(\rho \lambda_i)^k}{k}
\end{equation*}
to get
\begin{equation*}
\sum_{i=1}^n\ln(1+\rho \lambda_i) = \sum_{k=1}^{\infty}(-1)^{k-1}\frac{\rho^k}{k}\left(\sum_{i=1}^n\lambda_i^k\right).
\end{equation*}
Consequently, 
\begin{align}\label{ErgCap}
C(\rho) &= \sum_{k=1}^{\infty}\frac{(-1)^{k-1}}{k}\frac{\rho^k}{k!}\sum_{i=0}^{k-1}(-1)^i\binom{k-1}{i}\nonumber \\ & \times \prod_{j=-i}^{k-i-1}\frac{(n+j)(a+n+j-1)}{(a+b+2n+j-2)}.
\end{align}
Changing the summation order and performing the index change $k \mapsto k+i+1$ in \eqref{ErgCap}, we get
\begin{align*}
C(\rho) &= \sum_{i=0}^{\infty}(-1)^i\sum_{k = 0}^{\infty}\frac{(-1)^{k+i}}{(k+i+1)}\frac{\rho^{k+i+1}}{(k+i+1)!}\binom{k+i}{i} \nonumber \\ &  \prod_{j=-i}^{k}\frac{(n+j)(a+n+j-1)}{(a+b+2n+j-2)}.
\end{align*}
Now, one can observe that the product displayed in the right hand side of the last equality vanishes whenever $i \geq n$ due to the presence of the factor $j+n, -i \leq j \leq k$. Thus, the first series terminates at $i=n-1$ and together with the index change $j \mapsto n+j$ in the product lead to 
\begin{align*}
C(\rho) &= \sum_{i=0}^{n-1}\sum_{k = 0}^{\infty}\frac{(-1)^{k}}{(k+i+1)}\frac{\rho^{k+i+1}}{(k+i+1)!}\binom{k+i}{i} \\ &\prod_{j=n-i}^{k+n}\frac{(j)(a+j-1)}{(a+b+n+j-2)}.
\end{align*}
Next, we compute for each $n-i \leq j \leq n+k$ 
\begin{equation*}
\prod_{j=n-i}^{n+k}(j) = \frac{(n+k)!}{(n-i-1)!} = \frac{(n+1)_k n!}{(n-i-1)!},
\end{equation*}
and similarly
\begin{eqnarray*}
\prod_{j=n-i}^{n+k}(a+j-1) & = & \frac{(a+n)_k(a)_n}{(a)_{n-i-1}} \\
\prod_{j=n-i}^{n+k}(a+b+n+j-2) & = &  \frac{(a+b+n-1)_{n+k}}{(a+b+n-1)_{n-i-1}}
\end{eqnarray*}
Altogether, the ergodic capacity reads
\begin{align*}
\frac{(a)_n}{(a+b+n-1)_n}\sum_{i=0}^{n-1}\frac{n!}{(n-1-i)! i!}\frac{(a+b+n-1)_{n-i-1}}{(a)_{n-i-1}} \\ \sum_{k \geq 0}\frac{(-1)^k\rho^{k+i+1}}{(k+i+1)^2} \frac{(n+1)_k(a+n)_k}{(a+b+2n-1)_k k!}.
\end{align*}
But the series 
\begin{align*}
\sum_{k \geq 0}\frac{(-1)^k\rho^{k+i+1}}{(k+i+1)^2} \frac{(n+1)_k(a+n)_k}{(a+b+2n-1)_k k!}
\end{align*}
as well as its derivatives with respect to $\rho$ converge uniformly in any closed sub-interval in $]0,1[$. It follows that  
\begin{align*}
D_{\rho}(\rho D_{\rho})\sum_{k \geq 0}\frac{(-1)^k\rho^{k+i+1}}{(k+i+1)^2} \frac{(n+1)_k (a+n)_k}{(a+b+2n-1)_k k!} = \\  \rho^i{}_2F_1(n+1, a+n, a+b+2n-1; -\rho)
\end{align*}
where $D_{\rho}$ is the derivative operator acting on the variable $\rho$. Finally, the index change $i \mapsto n-i-1$ together with 
\begin{equation*}
(1-n)_i = (-1)^i\frac{(n-1)!}{(n-1-i)!}
\end{equation*}
yield 
\begin{align*}
\sum_{i=0}^{n-1}\frac{n!}{(n-1-i)! i!}\frac{(a+b+n-1)_{n-i-1}}{(a)_{n-i-1}}\rho^i = \\ \frac{n!\rho^{n-1}}{(a)_{n-1}} P_{n-1}^{a-1,b}\left(\frac{\rho+2}{\rho}\right) .
\end{align*}
Since 
\begin{equation*}
\frac{n!\rho^{n-1}}{(a)_{n-1}}\frac{(a)_n}{(a+b+n-1)_n} = \frac{n!(a+n-1)\rho^{n-1}}{(a+b+n-1)_n},
\end{equation*}
The statement of the proposition \ref{prop1} corresponds to the special parameters $a = r-t+1$ and $b = m-t-r+1$.

\section{Proof of Theorem 1}
\label{appen2}
Let's $n = t$ and $\rho \in [0,1]$. From \cite[Eq. (4.4.6)]{Ism}, we readily deduce that the hypergeometric function 
\begin{equation*}
{}_2F_1(n+1, a+n, a+b+2n-1; -\rho)
\end{equation*}
coincides up to a multiplicative factor with the Jacobi function of the second kind $Q_n^{a-1,b-2}$ in the variable $x$ related to $\rho$ by 
\begin{equation*}
- \rho = \frac{2}{1-x} \quad \Leftrightarrow \quad x = \frac{\rho+2}{\rho}.
\end{equation*}
Consequently, 
\begin{align*}
\left[D_{\rho}(\rho D_{\rho})\right] C(\rho) &= 2 B_{a,b,n} \frac{(1+\rho)^{b-2}}{\rho^{a+b-1}}P_{n-1}^{a-1,b}\left(\frac{\rho+2}{\rho}\right) \\ & Q_n^{a-1,b-2}\left(\frac{\rho+2}{\rho}\right)
\end{align*}
where 
\begin{equation*}
B_{a,b,n} = \frac{n!\Gamma(a+b+n-1)}{\Gamma(a+n-1)\Gamma(N+n-1)}.
\end{equation*}
Moreover, recall from \cite[Eq. (4.4.2)]{Ism}, that (note that $(\rho+2)/\rho > 1$)
\begin{align*}
Q_n^{a-1,b-2}\left(\frac{\rho+2}{\rho}\right) &= \frac{\rho^{a+b-3}}{2^{a+b-4}(\rho+1)^{b-2}}\int_{-1}^1(1-u)^{a-1} \nonumber \\ & \times (1+u)^{b-2}\frac{P_n^{a-1,b-2}(u)}{((\rho+2)/\rho) - u} du.
\end{align*}

After some mathematical manipulation and since 
\begin{equation*}
u \mapsto \frac{1}{((\rho+2)/\rho) - u}\left(P_{n-1}^{a-1,b}\left(\frac{\rho+2}{\rho}\right) - P_{n-1}^{a-1,b}(u)\right)
\end{equation*}
is a polynomial of degree $n-2$, then the orthogonality of the Jacobi polynomials entails
\begin{align*}
\left[D_{\rho}(\rho D_{\rho})\right] C(\rho) & =  \frac{B_{a,b,n}}{2^{a+b-3}}\int_{-1}^1(1-u)^{a-1}(1+u)^{b-2} \\ & \times P_{n-1}^{a-1,b}(u)\frac{P_n^{a-1,b-2}(u)}{\rho(\rho+2- \rho u)} du. 
\end{align*}
Writing 
\begin{equation*}
\frac{1}{\rho(\rho+2- \rho u)} = \frac{1}{2}\left[\frac{1}{\rho} - \frac{(1-u)}{\rho+2-\rho u}\right], \, u \in [-1,1],
\end{equation*}
and using again the orthogonality of Jacobi polynomials, we get 
\begin{align*}
\left[D_{\rho}(\rho D_{\rho})\right] C(\rho) &= -\frac{B_{a,b,n}}{2^{a+b-2}}\int_{-1}^1(1-u)^{a}(1+u)^{b-2} \\ & \times \frac{P_{n-1}^{a-1,b}(u)P_n^{a-1,b-2}(u)}{(\rho(1-u)+2)} du
\end{align*}
which makes sense for $\rho=0$. A first integration with respect to $\rho$ gives 
\begin{align*}
[\rho D_{\rho}]C(\rho) & = -\frac{B_{a,b,n}}{2^{a+b-2}}\int_{-1}^1(1-u)^{a-1}(1+u)^{b-2}P_{n-1}^{a-1,b}(u)\\ & \times P_n^{a-1,b-2}(u) [\ln(\rho(1-u)+2) - \ln 2] du
\end{align*}
and a second integration leads to 
\begin{align*}
C(\rho) & = -\frac{B_{a,b,n}}{2^{a+b-2}}\int_{-1}^1(1-u)^{a-1}(1+u)^{b-2}P_{n-1}^{a-1,b}(u) \\ & \times P_n^{a-1,b-2}(u) \left\{\int_0^{\rho} \frac{\ln(v(1-u)/2+1)}{v} dv\right\} du.
\end{align*}
Performing the variable changes $u \mapsto 1-2u$ in the last expression, we end up with
\begin{align*}
C(\rho) & = -B_{a,b,n}\int_{0}^1u^{a-1}(1-u)^{b-2}P_{n-1}^{a-1,b}(1-2u) \\ & \times P_n^{a-1,b-2}(1-2u)  \left\{\int_0^{\rho} \frac{\ln(vu+1)}{v} dv\right\} du
\end{align*}
for any $\rho \in [0,1[$. By analytic continuation, this formula extends to the cut plane $\mathbb{C} \setminus (-\infty, 0)$ and is in particular is valid for $\rho \geq 0$. Specializing it to $a = r-t+1$, and $b = m-t-r+1$  completes the proof of the Theorem 1.

\section{Proof of Theorem 2}
\label{appen3}
Perform the variable change $\rho \mapsto b\rho$ in the definition of $C_{t,r}^{m,\rho}$:  
\begin{align*}
C(b\rho) &= Z_{a,b,n}^{-1} \int \ln\left(\prod_{i=1}^n(1+b\rho\lambda_i)\right)\prod_{i=1}^n\lambda_i^{a-1}(1-\lambda_i)^{b-1} \\ & \times V(\lambda)^2{\bf 1}_{\{0 < \lambda_1 < \dots < \lambda_n < 1\}}d\lambda
\\& =  \frac{Z_{a,b,n}^{-1}}{b^{(an + n(n-1))}}  \int \ln\left(\prod_{i=1}^n(1+\rho\lambda_i)\right)\prod_{i=1}^n\lambda_i^{a-1} \\ & \times \left(1-\frac{\lambda_i}{b}\right)^{b-1} V(\lambda)^2{\bf 1}_{\{0 < \lambda_1 < \dots < \lambda_n < b\}}d\lambda.
\end{align*}
On the other hand, our obtained expression for the ergodic capacity together with  the variable change $ v \mapsto bv$ entail:
\begin{align*}
C(b\rho) &= -\frac{B_{a,b,n}}{b^a}\int_{0}^1u^{a-1}\left(1-\frac{u}{b}\right)^{b-2}P_{n-1}^{a-1,b}\left(1-\frac{2u}{b}\right)\\ & \times P_n^{a-1,b-2}\left(1-\frac{2u}{b}\right)   \left\{\int_0^{\rho} \frac{\ln(vu+1)}{v} dv\right\} du 
\end{align*}
Now 
\begin{equation*}
\lim_{b \rightarrow \infty} \frac{B_{a,b,n}}{b^a} = \frac{n!}{\Gamma(a+n-1)}
\end{equation*}
and similarly 
\begin{equation*}
\lim_{b \rightarrow \infty}\frac{Z_{a,b,n}^{-1}}{b^{n(a + n-1)}} = \prod_{i=1}^n\frac{1}{\Gamma(i)\Gamma(a+i-1)}
\end{equation*}
Moreover, the limiting transition \eqref{LimTra} yields
\begin{eqnarray*}
\lim_{b \rightarrow \infty} P_{n-1}^{a-1,b}\left(1-\frac{2u}{b}\right) &=&  L_{n-1}^{a-1}(u) \\ 
 \lim_{b \rightarrow \infty} P_n^{a-1,b-2}\left(1-\frac{2u}{b}\right) &=&  L_{n}^{a-1}(u).
\end{eqnarray*}
 As a result, 
\begin{align*}
\lim_{b \rightarrow \infty} C(b\rho) & = -\frac{n!}{\Gamma(a+n-1)}\int_{0}^{+\infty}u^{a-1}e^{-u}L_{n-1}^{a-1}(u)\\ & \times L_{n}^{a-1}(u)
 \left\{\int_0^{\rho} \frac{\ln(vu+1)}{v} dv\right\} du. 
 \end{align*}
where $\prod_{i=1}^n\frac{1}{\Gamma(i)\Gamma(a+i-1)}$ is the normalization constant of the density of the joint distribution of the ordered eigenvalues of a complex Wishart matrix \cite{P.J.Forrester_book}. The theorem is proved.

\section*{Acknowledgment}
The authors gratefully acknowledge Prof. M\'erouane Debbah for useful discussions and consultations, and R\'emi Bonnefoi who performed part of Matlab simulations.

\begin{IEEEbiography}[{\includegraphics[width=1in,height=1.25in,clip,keepaspectratio]{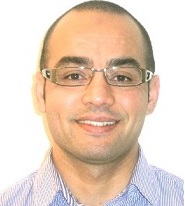}}]{Amor Nafkha} (S'03--M'08--SM'16) received the B.Sc. (Eng.) degree from the higher school of communications (SupCom), Tunis, Tunisia, in 2001, and the Ph.D. degrees from the University of South Brittany (UBS), Lorient, France, in 2006, all in Information and communications technology. From 2006 to 2007, he was a Postdoctoral researcher with the Signal, Communication, and Embedded Electronics (SCEE-IETR) research group at CentraleSup\'elec, France. During this time at SCEE, he was actively involved in the reconfigurable hardware platform implementation for software-defined radio, co-authoring several contributions on FPGA dynamic partial reconfiguration. Since January 2008, he has been an associate professor at CentraleSup\'elec, France. His research interests include multiuser and MIMO detection, hardware implementation, information theory, sample rate conversion, and spectrum sensing techniques. He has published more than 70 papers in international peer-reviewed journals and conferences.

\end{IEEEbiography}

\begin{IEEEbiography}[{\includegraphics[width=1in,height=1.25in,clip,keepaspectratio]{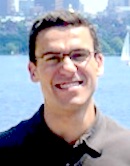}}]{Nizar Demni} recieved his BS on 2003 from University El Manar II in Tunisia and University of Aix-Marseille I in France. He obtained his MSC from University Paris VI at ’Laboratoire de Probabilites et modeles aleatoires’ and did his Ph.D. there under the supervision of Catherine Donati Martin. After the Ph.D. defense on 2007, Nizar Demni worked at Bielefeld university during two years with Professor Friedrich Goetze then worked at the engineering school of Bizerte during the year 2009-2010. Then, he obtained the position of associate Professor in Rennes 1 university and is up to now a member of the group ’Probabilites et theorie ergodique’. He defended his habilitation on December 2013 after significative research papers on Dunkl processes and free probability theory. He has published more than 40 papers in international peer-reviewed journals.

\end{IEEEbiography}

\EOD

\end{document}